\documentclass[graybox, envcountchap]{svmult}

\usepackage{mathptmx}        % selects Times Roman as basic font
\usepackage{helvet}          % selects Helvetica as sans-serif font
\usepackage{courier}         % selects Courier as typewriter font
\usepackage{makeidx}         % allows index generation
\usepackage{graphicx}        % standard LaTeX graphics tool
\usepackage{multicol}        % used for the two-column index
\usepackage[bottom]{footmisc}% places footnotes at page bottom

\makeindex             % used for the subject index
\usepackage{amsmath}
\usepackage{amssymb}
\usepackage{lscape}
\usepackage{multirow}

\def\gapp{\ifmmode\stackrel{>}{_{\sim}}\else$\stackrel{<}{_{\sim}}$\fi}
\def\gsim{\lower.5ex\hbox{\gtsima}}
\def\gtsima{$\; \buildrel > \over \sim \;$}
\def\lapp{\ifmmode\stackrel{<}{_{\sim}}\else$\stackrel{<}{_{\sim}}$\fi}
\def\lsim{\lower.5ex\hbox{\ltsima}}
\def\ltsima{$\; \buildrel < \over \sim \;$}

\newcommand\apgt{\ {\raise-.5ex\hbox{$\buildrel>\over\sim$}}\ }
\newcommand\aplt{\ {\raise-.5ex\hbox{$\buildrel<\over\sim$}}\ }
\begin{document}
\pagestyle{empty}
\frontmatter%%%%%%%%%%%%%%%%%%%%%%%%%%%%%%%%%%%%%%%%%%%%%%%%%%%%%%

\include{dedic}
\include{foreword}
\include{preface}

\mainmatter%%%%%%%%%%%%%%%%%%%%%%%%%%%%%%%%%%%%%%%%%%%%%%%%%%%%%%%

\setcounter{chapter}{8}

\title{Formation Channels for Blue Straggler Stars}

%\bigskip

\author{Melvyn B. Davies}

\institute{Melvyn B. Davies \at Department of Astronomy
and Theoretical Physics, Lund University, Sweden, \email{mbd@astro.lu.se}}

\maketitle
\label{Chapter:Davies}

\abstract*{In this chapter we consider two formation channels for blue straggler stars: 1) the merger of two single stars via a collision, and 2) those produced via mass transfer within a binary.
We review how computer simulations show that stellar collisions are likely to lead to relatively little mass loss and are thus effective in producing a young population of more-massive stars. The number of blue straggler stars produced by collisions will tend to increase with cluster mass. We review how the current population of blue straggler stars produced from primordial binaries decreases with increasing cluster mass. This is because exchange encounters with third, single stars in the most massive clusters tend to reduce the fraction of binaries containing a primary close to the current turn-off mass. Rather, their primaries tend to be somewhat more massive and have evolved off the main sequence, filling their Roche lobes in the past, often converting their secondaries into blue straggler stars (but more than 1 Gyr or so ago and thus they are no 
longer visible today as blue straggler stars).}

\section{Introduction}
\label{davsec:1}
%{\sl Set out review chapter. Explain sections and what people will see. Review
%three channels to make blues stragglers: 1) collisions/mergers; 2) mass transfer
%in binaries; and 3) via Kozai mechanism in triples. Perets \& Fabrycky 2009}

As has been discussed earlier in this book, blue stragglers sit on the main sequence
but above the current turn-off mass.
Their  existence is at first surprising: one would
have expected these stars to have evolved off the main sequence and become
white dwarfs\index{white dwarf} some time ago. How could a subset of stars somehow forget to
evolve off the main sequence\index{main sequence star}? In this chapter,
  we will focus on two, distinct, alternatives for blue straggler production: direct
collisions\index{collision} between two stars (leading to their merger), and mass transfer\index{mass transfer} (or merger\index{merger})
between two stars as part of natural evolution within a binary system.  We will see that both
formation mechanisms  probably occur, at least in some clusters\index{star cluster}. 
One should also note
that stellar mergers can occur also  in triples\index{triple system} as the inner binary is driven
into contact by the action of the third star via the Kozai effect\index{Kozai mechanism} \cite{PFa09}.
This pathway is discussed in detail in Chap. 11.

We begin by outlining some of the key concepts and ideas which will be
discussed in more detail in later sections of this chapter:

\begin{description}
\item {\bf a) Stellar collisions occur often 
in the cores of dense stellar clusters}\\
Physical collisions between stars
occur interestingly-often in the cores of the densest star clusters.  
Most of the collisions occur whilst the stars are on the main sequence\index{main sequence star}.
Because the relative speed of the stars within stellar clusters is much smaller than
their surface escape speeds, collisions between two main sequence stars
will lead to their merger with only a very small amount of mass loss.

\item {\bf b) The post-collision evolution of merger products is complex}\\
The post-collision evolution of merger products 
is complex, with many uncertainties. Merger products typically
contain a relatively large amount of angular momentum. In other words, the stars
will be rotating\index{rotating star} sufficiently rapidly that they may be significantly non-spherical.
Mixing\index{mixing} within the merger product is critical when one considers the refueling
of the core with unburned hydrogen and thus calculations of the
 lifetime of the merger product.
 
\item {\bf c) Encounters involving binary stars can be as frequent as those
involving only single stars}\\
A binary poses a much larger target for encounters than single stars.
Thus even if only a small fraction of stars are contained in binaries, the event
rate for strong encounters between binaries and single stars can be comparable to
(or possibly even in some cases exceed) the event rate for encounters between
two single stars.  Stellar collisions will occur in a subset of encounters involving binaries
thus potentially producing blue stragglers. Stellar collisions occurring during
encounters between two binaries will be important in less-dense clusters.

\item {\bf d) Blue stragglers may also be produced through the natural
evolution of isolated binaries}\\
In the case of very tight binaries, angular momentum loss\index{angular momentum loss} via winds\index{stellar wind}
may drive the two stars together forming a merger\index{merger} product not very different
from that formed via collisional mergers. Alternatively when the primary star
evolves, it may fill its Roche lobe\index{Roche lobe overflow} and transfer mass\index{mass transfer} on to the secondary star.
If the mass transfer is stable, this may add sufficient mass to the secondary to
convert it into a blue straggler.

\item {\bf e) The observed blue straggler population is probably a combination
of those formed via mergers and those formed from the evolution of binaries}\\
Blue stragglers can be formed either via mergers (as the outcome of collisions)
or via mass transfer (or merger) as an outcome of isolated-binary evolution.
 We will see that a combination of
both formation channels\index{formation channel} probably occurs in most globular clusters\index{globular cluster}, whilst binary
evolution\index{binary star
evolution} is most likely to be more important in less-dense clusters such
as open clusters\index{open cluster}, and in the halo\index{halo}.
\end{description}

This chapter is arranged as follows. In Sec.~\ref{davsec:2}, we review stellar collisions
considering the collision rate within clusters, which types of 
stars are likely to be involved in 
collisions, and their immediate outcome. We review the post-collision evolution
of merger products in Sec.~\ref{davsec:3}.
In Sec.~\ref{davsec:4}, we describe encounters between binary star systems and single
stars or other binaries. The evolution of isolated binaries is dealt with in Sec.~\ref{davsec:5}, where
we consider the effect of mass transfer on enhancing the mass of the secondary star,
potentially converting it into a blue straggler. In Sec.~\ref{davsec:6} we consider the blue straggler population which may be produced
as a combination of those formed via collisions and those formed via mass
transfer as part of the evolution of binaries.

%%%%%%%%%%%%%%%%%%%%%%%%%%%%%%%%%%%%%%%%%%%%%%%%%%
%%%%%%%%%%%%%%%%%%%%%%%%%%%%%%%%%%%%%%%%%%%%%%%%%%

%\newpage
\section{Stellar Collisions}
\label{davsec:2}
%{\sl Collision timescales: formula + explain gravitational focussing.
%What can collide with what. Zig-zag plot.
%Idea that $v_{inf}$ positive but small compared to $v_{esc}$ and why this matters.
%Outcomes of collisions. Idea of tidal capture.}

In order for collisions\index{collision} to contribute significantly to the observed blue straggler
population, we need a number of conditions to be satisfied: 1) collisions must
occur at interesting rates within stellar clusters\index{stellar cluster}; 2) collisions must lead to the merger\index{merger}
of stars; 3) the merger product must look like a moderately-massive main sequence
star (i.e. consistent with observations); and 4) the merger products must have a 
sufficiently long lifetime to produce a sufficiently large population of blue stragglers.
We consider points 1) and 2) in this section, and points 3) and 4) in Sec.~\ref{davsec:3}.

We begin by calculating the stellar encounter rate within clusters.
The cross section\index{cross section} for two stars, having a relative velocity at infinity
of $V_\infty$, to pass within a distance $R_{\rm min}$ is given by

\begin{equation}
\sigma = \pi R_{\rm min}^2 \left( 1 + {V^2 \over V_\infty^2} \right)\ ,
\end{equation}

\noindent where $V$ is the relative velocity of the two
stars at closest approach in a parabolic encounter ({\it i.e.\ }
$V^2 = 2 G (M_1 + M_2)/R_{\rm min}$, where $M_1$ and $M_2$ are the masses
of the two stars).
The second term is due to the attractive gravitational force, and
is referred to as gravitational focussing\index{gravitational focussing}.
In the regime where $V \ll V_\infty$ (as might be the case in
galactic nuclei\index{galaxy} with extremely high velocity dispersions), we recover
the result, $\sigma \propto R_{\rm min}^2$. However, if $V \gg V_\infty$
as will be the case in systems with low velocity dispersions, such as
globular\index{globular cluster} and open clusters\index{open cluster}, $\sigma \propto R_{\rm min}$. This will have consequences
for the relative frequency of collisions at various stages of stellar evolution as
will see below.

One may estimate the timescale for a given star to undergo an encounter
with another star,
$\tau_{\rm coll} = 1/n \sigma v$. For clusters with low velocity dispersions,
we thus obtain

\begin{equation}
\tau_{\rm coll} \simeq  10^{11} {\rm yr} \left( {10^5{\rm pc}^{-3} \over
n } \right) \left( { V_\infty \over 10{\rm km/s} } \right) 
\left( { R_\odot \over R_{\rm min} } \right) \left( { M_\odot \over
M } \right) \ ,
\end{equation}

\noindent where $n$ is the number density of single stars of mass $M$.
For an encounter between two single stars to be hydrodynamically
interesting, we typically require $R_{\rm min} \sim 3 R_\star$
for $V_\infty = 10$ km/s
(see for example, \cite{DBH91}).
We thus see that for typical globular clusters\index{globular cluster}, where $n \sim 10^5$ stars/pc$^3$,
up to 10\% of the stars in the cluster cores will have undergone a
collision at some point during the lifetime of the cluster.

Stars spend a large fraction of their life on the main sequence,
where helium is produced through fusion of hydrogen within their cores. 
Once the hydrogen fuel is exhausted, stars then evolve off the main sequence\index{main sequence star}
and move towards the red giant\index{red giant} branch where hydrogen fusion reactions occur
in a shell above a contracting helium core, as the surrounding envelope expands
to about 100 R$_\odot$.  For low-mass stars, once helium ignition occurs
within the core, the star shrinks to about 10 R$_\odot$ as it stays on the 
horizontal branch\index{horizontal branch}, fusing helium into carbon in its core. Post-horizontal
branch, the star evolves up the asymptotic giant branch, expanding to
length scales of around 300 R$_\odot$. Low-mass stars then eject their
envelopes producing a white dwarf\index{white dwarf}, whilst stars more massive than
8 M$_\odot$ will explode as a core-collapse supernova, leaving either
a neutron star\index{neutron star} or (for the most massive stars) a stellar mass black hole\index{black hole} (see Chap. 1 for a more detailled discussion).

\begin{figure}[t]
\sidecaption
%\centering
%\resizebox{11truecm}{!}{\includegraphics{isothermal.eps}}
\includegraphics[width=75mm]{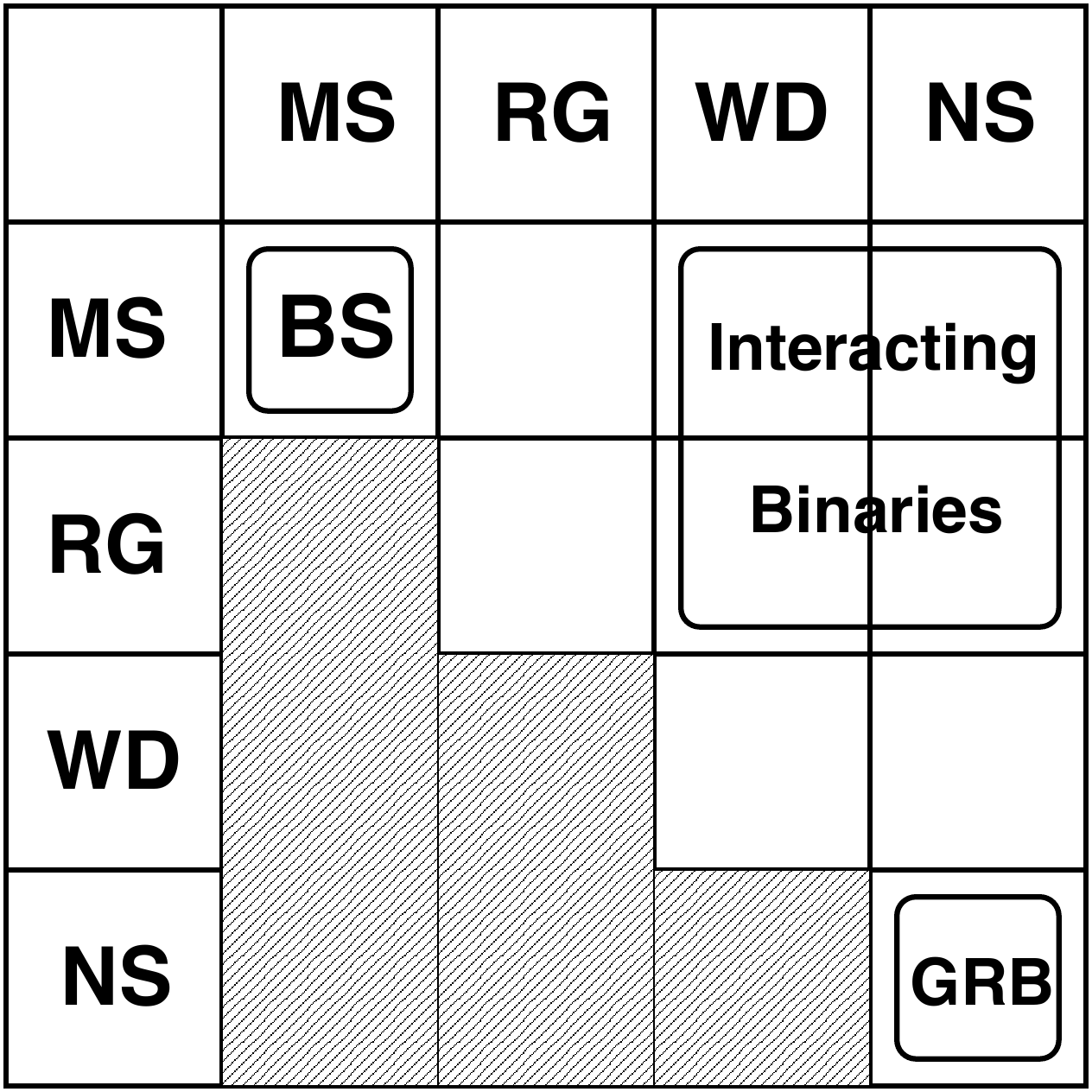}
%\begin{minipage}{13truecm}
\caption{Plot showing the grid of possible collisions between various 
stellar species: main sequence stars\index{main sequence star} (MS), red giants\index{red giant} (RG), white dwarfs\index{white dwarf} (WD)
and neutron stars\index{neutron star} (NS). Collisions between two main sequence stars may produce
at least some of the observed blue stragglers (BS). Collisions between either
main sequence stars or red giants and white dwarfs or neutron stars may produce
interacting binaries (cataclysmic variables\index{cataclysmic variable} and low-mass X-ray binaries\index{low-mass X-ray binary}). Encounters
involving two neutron stars could potentially produce gamma-ray bursts\index{gamma-ray burst} (GRB).}
\label{mdavies_figure1}
%\end{minipage}
\end{figure}

It is possible for stars to be involved in collisions during all of the phases
of stellar evolution described above, as shown in Fig.~ \ref{mdavies_figure1}. 
Of interest to us here are the collisions involving two main sequence stars which may
produce at least some of the observed blue stragglers.

We consider now  when stars are most likely to be involved
in collisions\index{collision}. One may integrate the collision rate equation over the entire
lifetime of a cluster to to calculate the expected number of collisions
 $n_{\rm coll}$ for a particular star:

\begin{equation}
n_{\rm coll}(t) = \int_0^t \Gamma_{\rm coll} dt = n \int_0^t  \sigma(R_{\star}) V_\infty dt \ ,
\end{equation}
where $\Gamma_{\rm coll}$ is the collision rate for the star and $\sigma(R_{\star})$
 is the collision cross section\index{cross section} (as given in Eq.~(1) with the minimum 
 distance $R_{\rm min}$ set to the stellar radius\index{stellar radius} $R_\star$) which will change
as a function of time as the star evolves (and its radius changes).

\begin{figure}
\sidecaption
%\centering
%\resizebox{11truecm}{!}{\includegraphics{isothermal.eps}}
\includegraphics[width=75mm]{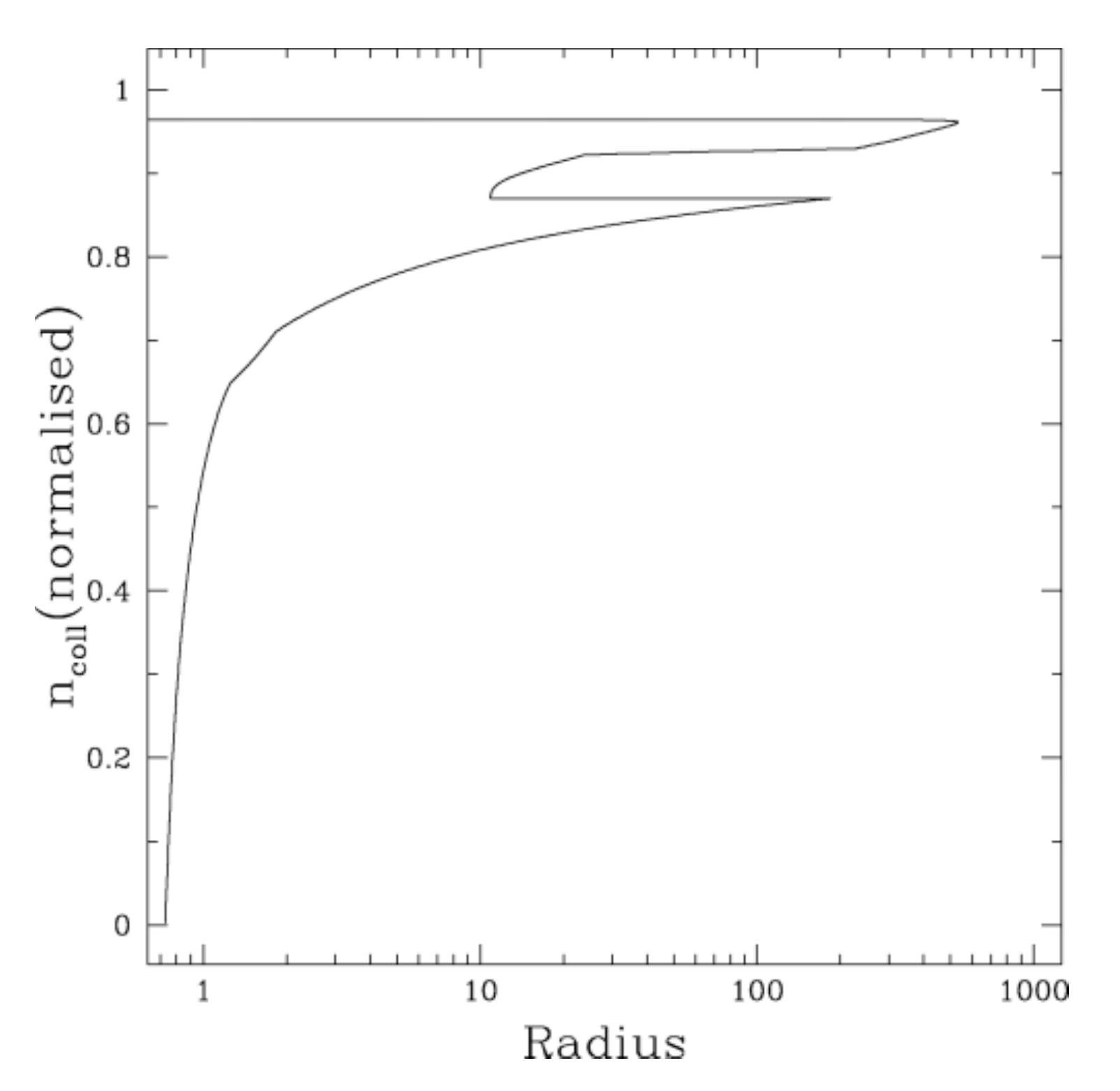}
%\begin{minipage}{13truecm}
\caption{The cumulative number of collisions as a function of stellar
radius (in solar units)  for 0.4 M$_\odot$ mass stars in a globular cluster.
The expected number of collisions has been normalised so that the total
number of collisions over the entire cluster lifetime is one. The various
phases of stellar evolution\index{stellar evolution} are clear from this plot: the main sequence
phase ending when the stellar radius is few solar radii, the red giant
phase extending up to a radius $\sim 200$ R$_\odot$, the star has
a radius $\sim 10-20$ R$_\odot $ on  the horizontal branch\index{horizontal branch}, then expands
again to over 300 R$_\odot$ on the asymptotic giant branch\index{asymptotic giant branch star}. The results
obtained for 0.6 and 0.8 M$_\odot$ models are very similar.
(Figure 1 from \cite{SAD05}, reproduced with permission)}
\label{mdavies_figure2}
%\end{minipage}
\end{figure}

The number density of stars in a cluster $n$ is assumed to be constant throughout
the evolution of the star. The result of such a calculation is show in 
Fig.~\ref{mdavies_figure2}, where we plot the normalised expected number
of collisions as a function of stellar radius for 0.4 M$_{\odot}$ stars in a globular cluster\index{globular cluster} 
(stellar radius is used here as it easily shows the various phases of stellar
evolution). The frequency of collisions when the star is large are somewhat
reduced because of the effects of gravitational focussing\index{gravitational focussing} as the cross
section $\sigma(R_{\star}) \propto R_{\star}$ rather than $R_{\star}^2$.
From Fig.~\ref{mdavies_figure2}, we can see that some 
60 -- 70\% of collisions will occur whilst the 
star is on the main sequence, a further 20\% or so occur whilst the star
is ascending the giant branch, 10\% whilst the star is on the horizontal
branch and a little less than 10\% on the asymptotic giant branch. We thus
conclude that the majority of collisions will occur whilst the star is on the main 
sequence.

We consider now the immediate outcomes of collisions\index{collision} between two main sequence stars\index{main sequence star}.
Such collisions are complex events. Understanding them well requires 
fully 3D computational hydrodynamic simulations\index{hydrodynamic simulation}. Much work has been
done modelling such collisions, particularly involving low-mass main sequence 
stars with
relatively-low velocities, which are relevant for the encounters of interest to us
in globular clusters (including \cite{davBH87,LRS95,LRS96,Lom02,SADB02}). 

There are two speeds to consider in a stellar collision:
the relative speed of the two stars at infinity $V_\infty$ and the surface escape speeds
of the stars ($V_{\rm esc} = \sqrt{2 G M_\star / R_\star}$). For globular clusters\index{globular cluster},
$V_\infty \simeq 10$ km/s. In comparison, for low-mass main sequence stars,
$V_{\rm esc} \simeq 600$ km/s. We should not be surprised therefore to find
that collisions in globular clusters lead to mergers having little mass loss
(typically 1--10 \% of mass is lost; e.g., \cite{davBH87,davBH92})
as the ejection of a small fraction of the total mass can carry off the (small)
positive energy contained in the collision. 

\begin{figure}
%\centering
%\resizebox{13truecm}{!}{
\includegraphics[width=119mm]{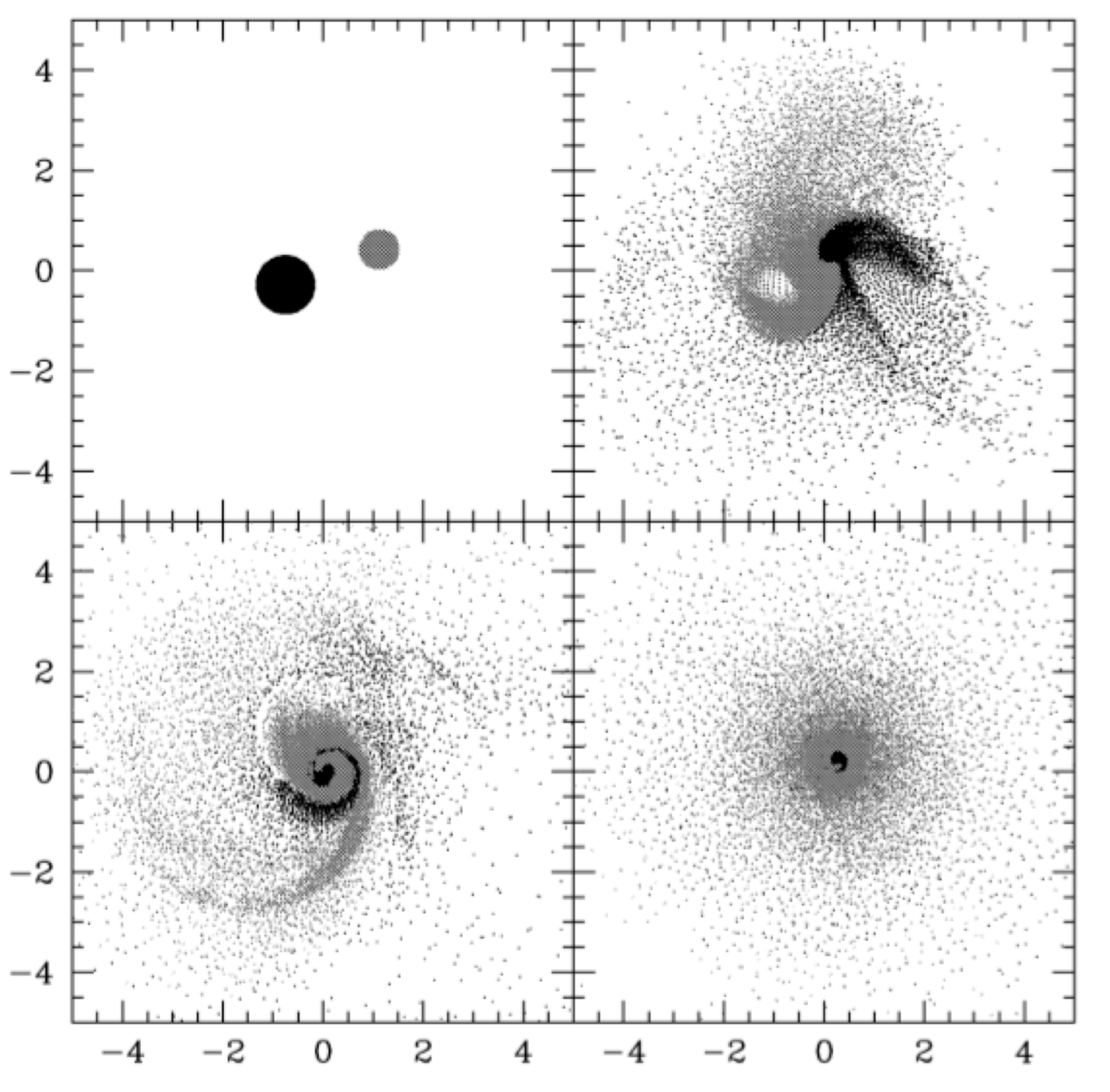}
%\begin{minipage}{13truecm}
\caption{Snapshots of a collision between a 0.6 M$_\odot$ 
main sequence star (black dots) and a  0.4 M$_\odot$ 
main sequence star (grey dots).
The minimum distance between the stars in the initial collision was 
0.255 R$_\odot$, equal to $0.25 ( R_1 + R_2)$
and the stars had a relative speed at infinity $V_\infty = 10$ km/s.
The colours represent the density of the gas in the plane of the encounter.
(Figure 2 from \cite{SAD05}, reproduced with permission).}
\label{mdavies_figure3}
%\end{minipage}
\end{figure}

Snapshots of a typical collision between two low-mass main sequence stars
is shown in Fig.~ \ref{mdavies_figure3}. The stars quickly merge, with little mass
loss, although the merged object does contain (unsurprisingly) considerable
angular momentum\index{angular momentum}. The post-collision evolution of such a merged object is complex,
and will be discussed in the next section.

%\newpage

The energy lost in a head-on impact\index{head-on impact}
is equivalent to $\delta V_\infty \sim 100$Êkm/s. Stars will become bound
even for close encounters which do not (initially) lead to physical collisions\index{collision},
with the minimum distance to capture $R_{\rm capt} \sim 3 R_\star$. Indeed,
such a capture mechanism has been invoked as a way to produce the population
of low-mass X-ray binaries\index{low-mass X-ray binary} in globular clusters\index{globular cluster}, where in this case a passing 
neutron star\index{neutron star} captures a main sequence star via tidal interactions\index{tidal interaction} \cite{davFPR75}.

%%%%%%%%%%%%%%%%%%%%%%%%%%%%%%%%%%%%%%%%%%%%%%%%%%
%%%%%%%%%%%%%%%%%%%%%%%%%%%%%%%%%%%%%%%%%%%%%%%%%%

\section{Post-collision Evolution}
\label{davsec:3}

The subsequent evolution of the merger\index{merger} product is complicated,
though much modelling work has been done 
 --- see, for example, \cite{GP08,GPH08,SanBH97,SAD05,davSills01,SKL09,davSills97}.
%Glebbeek et al 2008; Glebbeek \& Pols 2008)

Even though most of the material is bound in a single merged object,
it will not immediately appear as a main sequence star\index{main sequence star}. The incoming
kinetic energy\index{kinetic energy} of the impact has been converted into thermal energy\index{thermal energy}, the
merger product is out of virial equilibrium\index{virial equilibrium} and will expand. 

One is interested to learn about the extent of the expansion and the timescales
involved for it to return to the main sequence, if indeed it does so.
The object typically contains significant angular momentum\index{angular momentum} as most collisions are
relatively grazing so the merger product contains the angular momentum from
the trajectories of the two stars. Indeed many merger products
initially contain too much angular momentum to contract down
to the main sequence.

By considering angular momentum loss through either disc- or wind-locking, it was shown that both methods allow the merger product to shed
sufficient angular momentum to contract down to the main sequence~\cite{SAD05}.

\begin{figure}
\sidecaption
%\centering
%\resizebox{11truecm}{!}{\includegraphics{isothermal.eps}}
%\resizebox{12truecm}{!}{
\includegraphics[width=75mm]{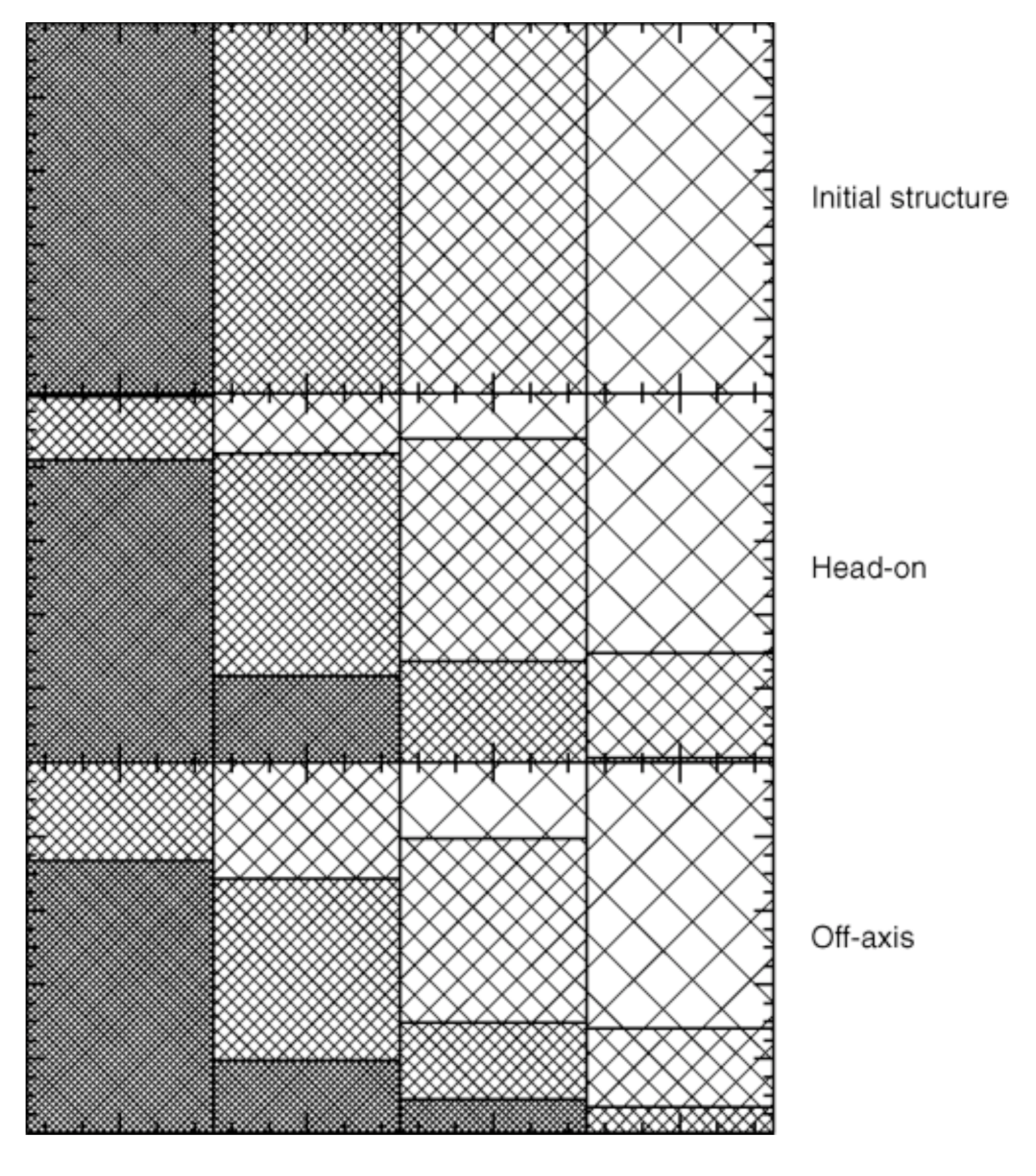}
%\begin{minipage}{13truecm}
\caption{A plot showing the mixing\index{mixing} of material which occurs
in head-on and  off-axis collisions\index{collision}
between two low-mass main sequence stars. The four columns
in each plot represent four equally-spaced mass bins. 
(Figure 3 from \cite{SAD05}, reproduced with permission)}
\label{mdavies_figure4}
%\end{minipage}
\end{figure}

In modelling the evolution, one is also concerned about how 
much mixing\index{mixing} takes place within the star.  Will the core be refueled with
a fresh supply of hydrogen\index{hydrogen} to thus extend the life of the blue straggler?

The distribution of matter in low-mass main sequence stars in illustrated in 
 Fig.~ \ref{mdavies_figure4} where we show the redistribution of material
 within the stars as an outcome of both head-on\index{head-on collision} and off-axis collisions~\cite{SAD05}\index{off-axis collision}.

If collisions were to lead to the complete mixing of material, then we would 
see that the four columns of the initial structure would be completely
mixed in the collision products. In other words, the inner mass quartile 
of the collision\index{collision} product would contain equal amounts of material from each
of the four mass quartiles from the initial structure. This is not seen. 

Indeed,
only a very small amount of material in the central mass quartile of the collision
products is drawn from outer regions of the pre-collision stars (very little mixing
is seen also in the work of \cite{SanBH97,davSills97}). Slightly more mixing\index{mixing} is seen in off-axis collisions than in head-on collisions
(see also \cite{davSills01}).  

Mixing is  important
when one considers the lifetime of the collision product: collisions involving
main sequence stars close to the turn-off mass, will contain relatively small
amounts of unburned hydrogen in their cores. The lifetime of the collision 
products will be short unless fresh hydrogen\index{hydrogen} can be brought in to the core
of the merger product. Indeed, recent work suggests that blue stragglers
have somewhat shorter lifetimes than regular main sequence stars of 
similar masses \cite{SGCR13}.

%\newpage

\begin{figure}[b]
\sidecaption
%\centering
%\resizebox{11truecm}{!}{\includegraphics{isothermal.eps}}
%\resizebox{13truecm}{!}{
%\includegraphics[width=119mm]{mdafig5}
\includegraphics[width=75mm]{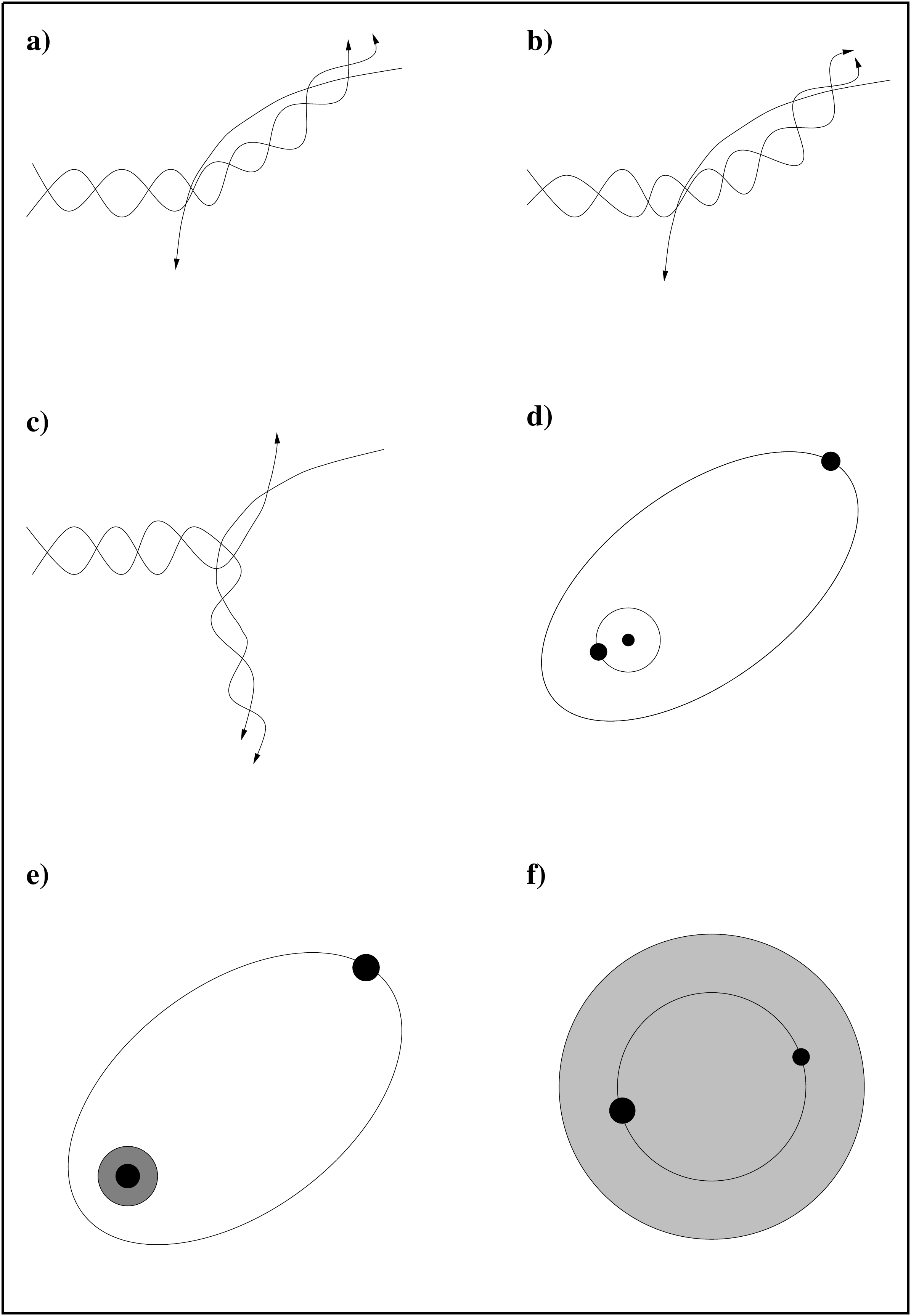}
%\begin{minipage}{13truecm}
\caption{Possible outcomes of encounters\index{encounter} between a binary and a single
star: a) a fly-by\index{fly-by} occurs where the binary's orbit is changed, b) the fly-by
leads to the merger of the two stars in the binary, c) the intruding star exchanges
into the binary, d) the system forms a (transient) triple system\index{triple system}, e) two of the stars
merge\index{merger} and remain bound to the third star, and f) a common envelope\index{common envelope} system is formed
where two of the stars orbit inside a gaseous envelope made from the third star.}
\label{mdavies_figure5}
%\end{minipage}
\end{figure}

\section{Encounters Involving Binary Stars}
\label{davsec:4}
%{\sl Cross sections. Possible outcomes. Idea that $\Gamma_{1+1} \sim  \Gamma_{2+1}$.
%Hard-soft boundary. Heggie's law. Thermal distribution of eccentricities.
%Clean exchanges changing mass within binaries. Stellar collisions during encounters.
%2+2 compared to 2+1. Cite work on 2+1 and 2+2 cross sections. Merged binaries 
%(MBD definition).
%Using up binaries in stellar clusters.}

In this section we consider the role played by encounters\index{collision}\index{encounter} involving binary
stars\index{binary
star} in producing blue stragglers. Binaries are larger targets than single
stars within stellar clusters. 
Interesting encounters occur when a star passes within a distance roughly
equal to the size of the binary. Hence, the time scale for encounters 
is given by 

\begin{equation}
\tau_{2+1} \simeq  10^{11} {\rm yr} \left( {10^5pc^{-3} \over
n } \right) \left( { V_\infty \over 10km/s } \right) 
\left( { R_\odot \over a_{\rm bin} } \right) \left( { M_\odot \over
M } \right) \ .
\end{equation}

One can see how encounters between binaries and single stars
can be as frequent as encounters between two single stars.
For example, if a cluster possesses a binary fraction\index{binary fraction} of around
0.05, then the encounter rates will be similar for binaries
of separation $a_{\rm bin} \sim$ 60 R$_\odot$. There 
are many outcomes possible for encounters involving binaries
as illustrated in Fig.~ \ref{mdavies_figure5}: fly-bys occur
where the binary retains its stellar components although the binding
energy\index{binding energy} and eccentricity\index{eccentricity} of the binary orbit may change; fly-bys may
lead to the merger of the two stars within the binary; an intruder
star may exchange into the binary with typically the least massive
of the three stars being ejected; the system may form a (transient) 
triple system\index{triple system}; two of the stars may merge\index{merger} but remain bound to the third
star; or a common envelope\index{common envelope} system may form where two of the stars
orbit inside a gaseous envelope made from the third star.

For us here, considering blue straggler production in stellar clusters,
we are particularly interested in the fifth possible outcome, where
two stars merge. The outcomes for such mergers are likely to be 
similar to those seen for collisions between two single stars as described
earlier. 

We will now consider  some general concepts concerning binary-single\index{binary star} 
encounters\index{encounter}. If binaries are sufficiently wide, encounters with single
stars will tend to break them up as the kinetic energy\index{kinetic energy} contained in the
incoming star exceeds the binding energy of the binary. Such binaries
are referred to in the literature as {\em soft}\index{soft binary}. Whereas binaries
which are more tightly bound and are thus resilient to break-up
are known as {\em hard}\index{hard binary}.  The separation of the stars in a binary
sitting on the hard-soft boundary depends on the masses of the stars
in the binary, and the mass of the incoming star. Assuming that
all stars are of one solar mass, the binary separation for a system
on the hard-soft boundary is given approximately by $a_{\rm hs}
\simeq 6 {\rm AU} (V_\infty / 10 {\rm kms^{-1}})^{-2}$.
Encounters tend to break up soft binaries, whereas hard binaries
get harder (i.e. more bound). They are also left with a thermal
distribution of eccentricities\index{eccentricity}, where the distribution follows
$dn/de = 2e$. As stated earlier, in exchange encounters
it is most often the least massive of the three stars which is ejected.

Thus, encounters involving binaries will tend to increase the mass
of the stellar components within binaries; a fact which will become
important later in this chapter when we compare the blue straggler
formation rates in clusters produced from the evolution of
 primordial binaries to the rate due to collisions and mergers.

\begin{figure}[b]
%\centering
%\resizebox{11truecm}{!}{\includegraphics{isothermal.eps}}
%\resizebox{10truecm}{!}{
\includegraphics[width=119mm]{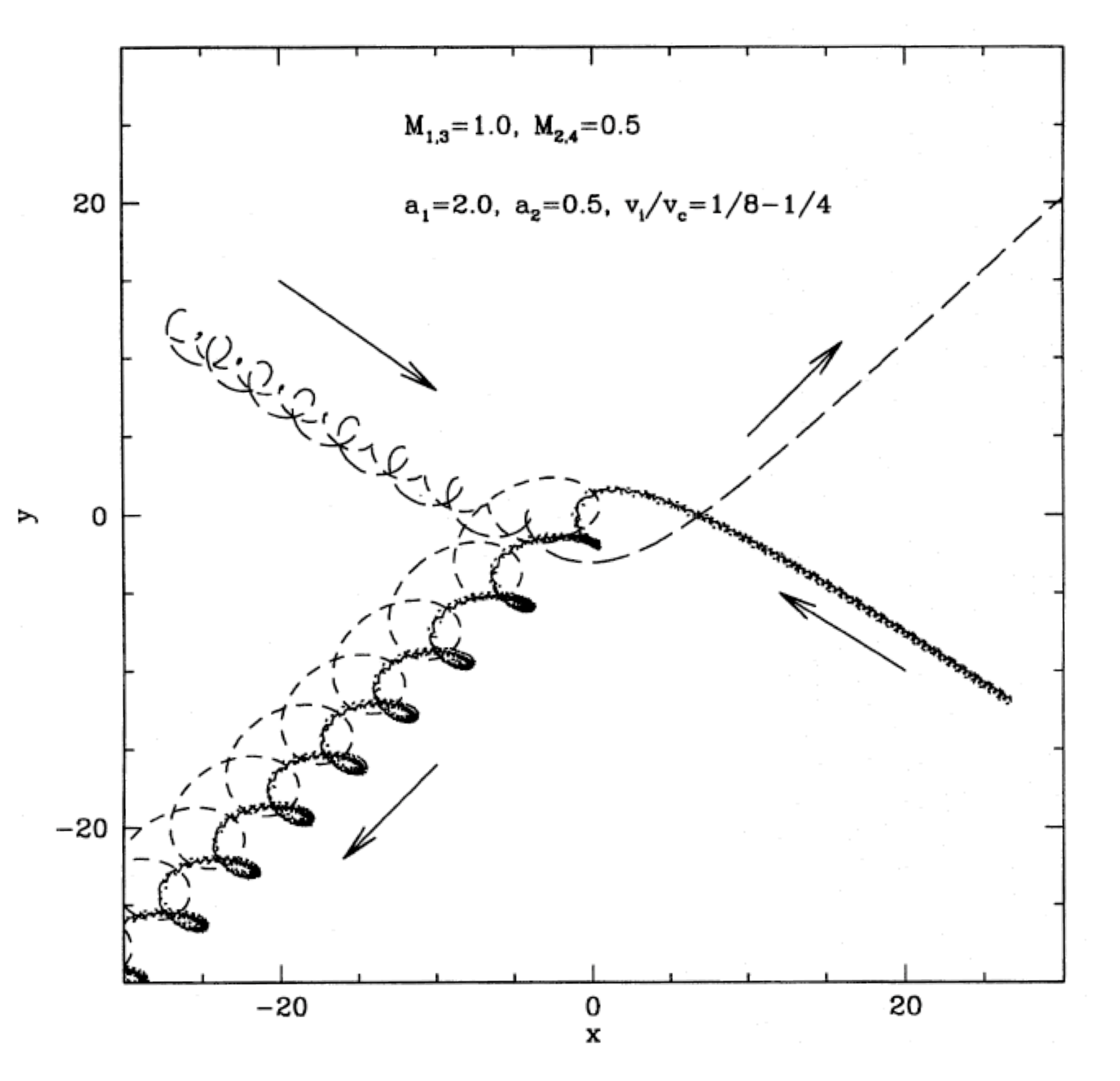}
%\begin{minipage}{13truecm}
\caption{An example of a binary-binary encounter\index{encounter}, in this case producing
a hierarchical triple\index{hierarchical triple system} and a single star.
(Figure 3 from \cite{Bacon96}, reproduced with permission)}
\label{mdavies_figure6}
%\end{minipage}
\end{figure}

The fraction of binary-single encounters which lead to collisions\index{collision} between
stars is a function of the binary separation. For binaries having
separations around 1 AU, the fraction of strong binary-single  encounters
where two stars pass within some distance $r_{\rm min}$ is
found, through numerical experimentation, to be $f  \propto 
(r_{\rm min}/a_{\rm bin})^\gamma$
where $\gamma \simeq 1/2$ \cite{DBH93,DBH94}. So, for example,
collisions and mergers occur in 10-20 \% of encounters involving solar-like stars\index{solar-type star}
and a binary of separation 1 AU.  Thus stellar mergers\index{merger} occurring during a binary-single
encounter\index{encounter} may make a significant contribution to the total merger rate within
a stellar cluster providing the binary fraction\index{binary fraction} is large. In typical globular clusters\index{globular cluster}, where
the binary fraction\index{binary fraction} is perhaps around 10 \% or smaller (e.g., \cite{Milone12}),
the collision\index{collision} rate derived from single-single collisions is likely to exceed that derived
from encounters involving binaries.

We now consider encounters involving {\em two} binaries\index{binary star}. The cross section\index{cross section}
for some kind of strong interaction for binary-binary encounters is in fact roughly
the same as for binary-single encounters: we require that the binaries pass within
a distance roughly comparable to the size of the binaries. However the fraction
of strong encounters\index{encounter} leading to physical collisions is larger. This can be seen
simply by reflecting that when we have four bodies (i.e. two binaries)
involved in a complex encounter
the number of distinct pairs $n_4 = 4 (4-1)/2 = 6$ whereas for three bodies
(i.e. a binary and a single star), the number of pairs is $n_3 = 3(3-1)/2 = 3$.

Thus, we have a much greater chance that at least one pair will suffer a close
passage during the whole encounter. Typically a binary-binary encounter quickly
resolves itself into a (transient or stable) triple and a single star. Such an encounter
is shown in Fig.~ \ref{mdavies_figure6}.

In sparse clusters, with lower number densities
of stars, collisions between two single stars may be rare. In such cases, collisions
occurring during the interaction between two binaries may dominate.
There are several papers which contain calculations of 
cross sections\index{cross section} for various outcomes (binary break-ups, exchange encounters,
and stellar collisions): including \cite{Bacon96,DBH93,DBH94,Heggie75,HMG92,HV83,davLeonard89,SP93}.

\begin{figure}[h]
%\centering
%\resizebox{11truecm}{!}{\includegraphics{isothermal.eps}}
%\resizebox{14truecm}{!}{
\includegraphics[width=119mm]{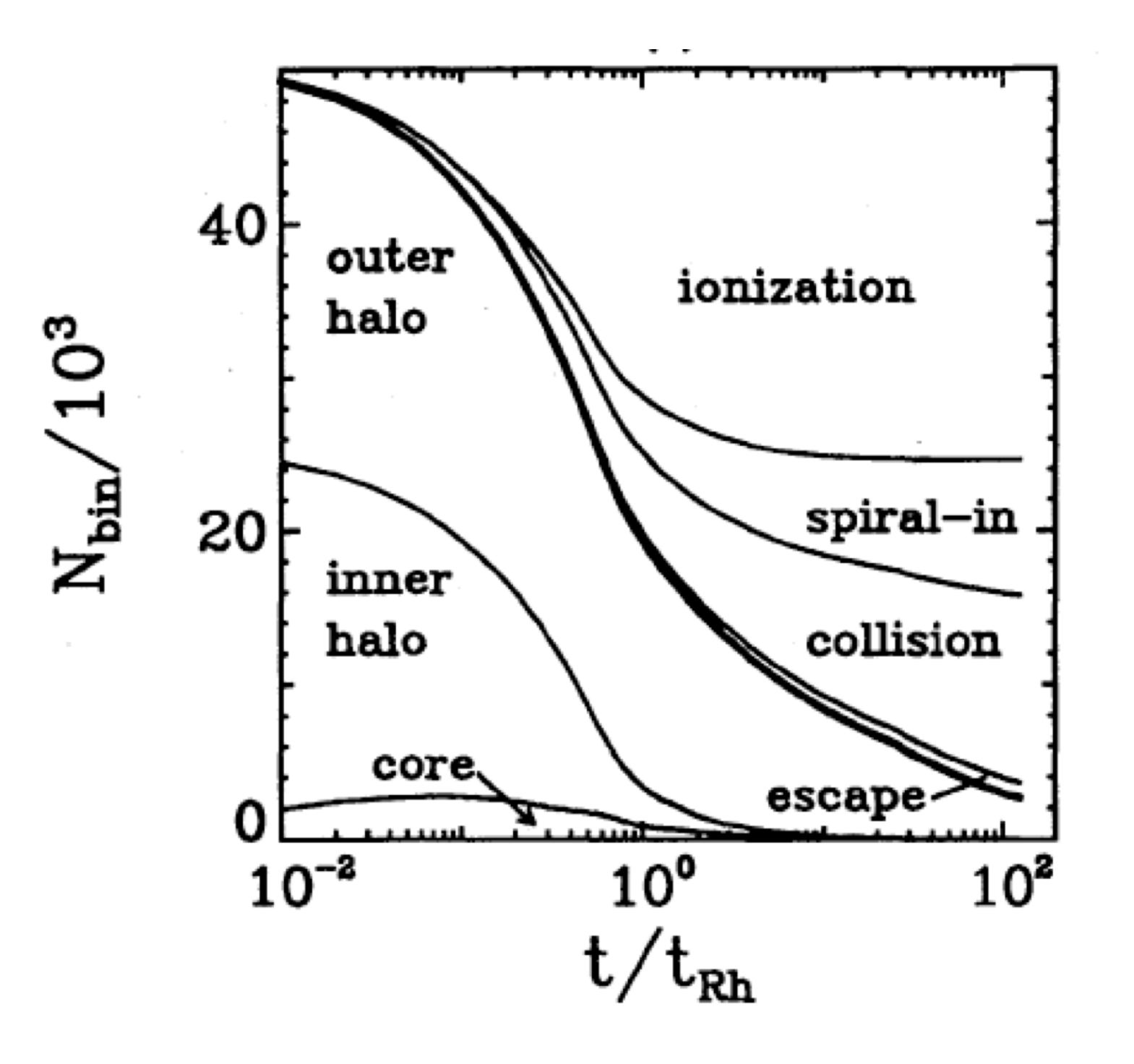}
%\begin{minipage}{13truecm}
\caption{The location and fate of binaries within a stellar cluster\index{stellar cluster}, containing initially
fifty thousand binaries\index{binary star}. The left-hand side of the figure shows the location
of any surviving binaries as a function of time (in units of the half-mass relaxation
time).  The right-hand side of the figure shows the fate of those binaries
which have not survived: break-up (ionisation), stars merging (spiral-in and collision),
and those (few) escaping from the cluster.
(Figure 3 from \cite{HMR92}, reproduced with permission)}
\label{mdavies_figure7}
%\end{minipage}
\end{figure}

Binaries are, in many ways, a fossil fuel of globular clusters\index{globular cluster}. Through close
encounters with single stars or other binaries, they are broken up or ground 
down, processed, and sometimes ejected from clusters. Those not formed in 
clusters' cores will sink into the core (as they are heavier than the average star) and
there undergo close encounters. The fate of binaries within a cluster is shown in 
Fig.~ \ref{mdavies_figure7}, taken from \cite{HMR92}, who used a simple
model to follow a population of primordial binaries within a cluster, allowing
them to sink into the cluster core, suffer encounters, be ejected from the core,
or the entire cluster. They found that about half of the binaries will be broken
up (often termed ``ionisation\index{ionisation}'') --- in fact often by encounters with other binaries.
Others will be involved in collisions\index{collision} or would be hardened\index{hard binary} to the point where
the two stars in the binary would merge\index{merger}. A very small fraction will escape
from the cluster.

\section{Making Blue Stragglers Via Binary Evolution}
\label{davsec:5}

In this section we consider the formation of blue stragglers through the evolution of
stellar binaries\index{binary star evolution}, where either the two stars spiral together and merge as angular
momentum\index{angular
momentum} is lost via stellar winds\index{stellar wind}, or where mass transfer\index{mass transfer} occurs from one (evolved)
star to the other. Both processes may produce stars more massive than the current
turn-off\index{turn-off}, providing in the first case the total mass of the two stars exceeds the current
turn-off, and in the second case providing the mass transferred from the primary to
the secondary increases the mass of the secondary above that of the turn-off mass
of the cluster.

\begin{figure}
%\centering
%\resizebox{11truecm}{!}{\includegraphics{isothermal.eps}}
%\resizebox{8truecm}{!}{
\includegraphics[width=119mm]{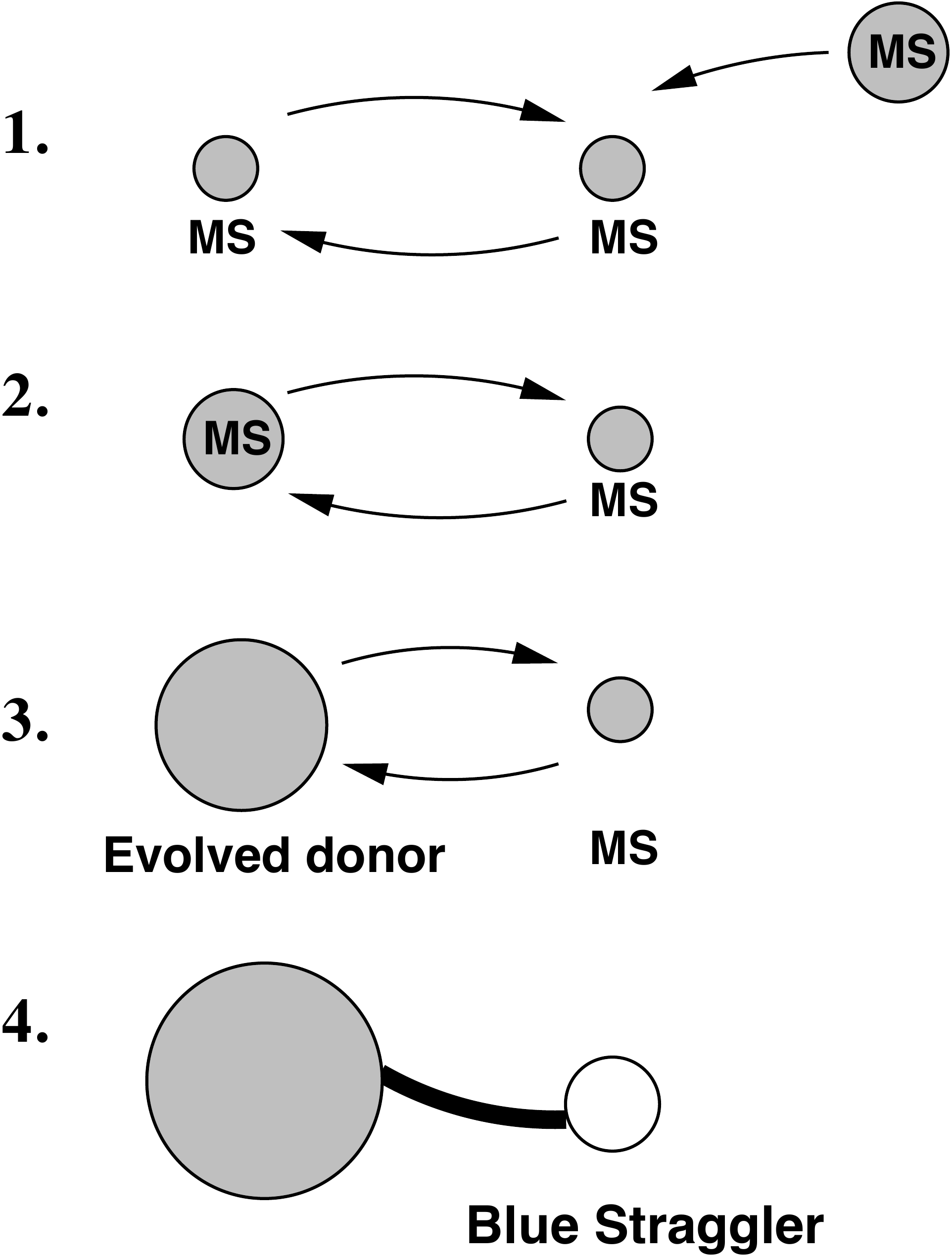}
%\begin{minipage}{13truecm}
\caption{The evolutionary pathway to produce blue straggler stars (BSSs) through mass transfer\index{mass transfer} in wide binaries\index{binary star} in globular clusters\index{globular cluster}. A more massive main sequence star exchanges into a binary containing 
two main sequence stars\index{main sequence star} (phase 1). The typical primary mass after encounters\index{encounter} in a 
sufficiently crowded cluster is M$_1 \simeq$1.5 -- 3M$_\odot$ \cite{DH98}. 
This primary 
evolves off the main sequence and fills its Roche lobe\index{Roche lobe overflow} (phase 3). The secondary gains mass 
from the primary becoming a BSS (phase 4) at a time roughly equal to the main sequence 
lifetime of the donor star. Hence BSSs have formed earlier in binaries containing 
more-massive primaries (i.e. in high collision rate clusters). Given the finite lifetime of BSSs, 
the BSS population in the most crowded clusters today could be lower than in very sparse 
clusters.
(Figure 4  from \cite{davDPd04}, reproduced with permission).}
\label{mdavies_figure8}
%\end{minipage}
\end{figure}

Tight binaries, where the separation is only a few times larger
than main sequence stars\index{main sequence star}, may merge as angular momentum\index{angular momentum loss} is lost
via stellar winds\index{stellar wind} \cite{Vilhu82}. The subsequent evolution of the merger\index{merger} product is
likely to be rather similar to that described earlier for objects produced
via stellar collisions\index{collision}. Here, as before, the object is likely to be spinning 
rapidly, perhaps  leading to global circulation within the merged object
which may help refuel the core with unburnt hydrogen.  Clearly the merger
product will be a single star, unless the tight binary is itself a component
of a wider binary (see \cite{PFa09} and Chap. 11).

We consider now the evolution of binaries which are too wide to merge
via angular momentum loss\index{angular momentum loss} from stellar winds\index{stellar wind}. In such systems, mass may flow
from the primary to the secondary star when the former evolves off the main sequence,
expanding up the giant branch and filling  its so-called Roche lobe\index{Roche lobe overflow} where
material at the primary's surface flows toward the secondary star
\cite{davMcCrea64}. This mass transfer\index{mass transfer}
may be stable, in the sense that the mass transfer rate does not grow rapidly,
 and the system
evolves steadily as the primary evolves up the giant branch\index{red giant} (as illustrated
schematically in Fig.~ \ref{mdavies_figure8}). In such a case, the secondary
will gain mass from the primary. 

 Alternatively, the
mass transfer can be unstable: the rate increases to the point where a very rapid
mass transfer occurs with a large fraction of the primary's envelope engulfing the 
secondary forming what is know as a common envelope\index{common envelope} system where the 
core of the primary and the secondary star orbit inside this envelope of gas. 
In this case, the core of the primary and the secondary star will spiral
together, dumping energy and angular momentum into the surrounding envelope
which will be ejected.

In order to compute whether mass transfer is unstable, one has to consider
the response of the donor star to its mass loss and compare this to how the
size of the Roche lobe changes as mass is transferred (see Chap. 8 for a more detailed discussion). Mass transfer will be 
unstable when the ratio of the donor radius to the Roche lobe radius increases,
in other words, when the star overfills its Roche lobe\index{Roche lobe overflow} by increasing amounts.

If the mass transfer
is {\it conservative}\index{conservative mass transfer}, meaning that mass transfers from one star to the other
without any loss of material (or angular momentum\index{angular momentum loss}) via stellar winds, 
then the separation of a binary will increase if the donor is less-massive
than the receiving star, and decrease if the donor is more-massive than the 
receiving star.  One can see that systems containing more massive donors
will often be unstable, as once filling their Roche lobes, the donors
will overflow their Roche lobes more and more, leading to extremely high
rates of mass transfer.

\begin{figure}[!b]
\sidecaption
%\centering
%\resizebox{9truecm}{!}{
\includegraphics[width=75mm]{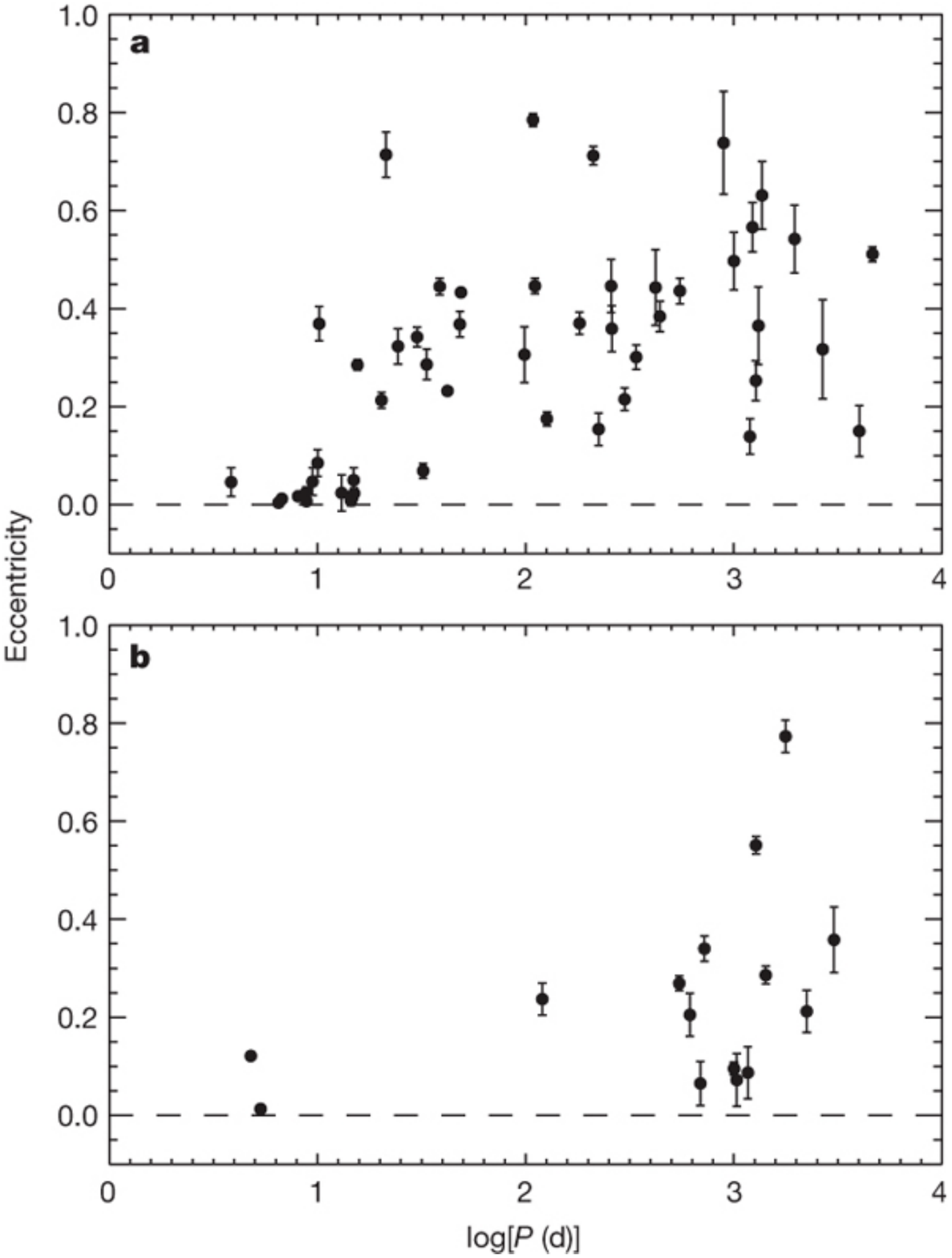}
%\begin{minipage}{13truecm}
\caption{Distribution of solar-type main sequence binaries\index{binary star} seen in the open cluster NGC 188\index{NGC 188} (upper figure), and the binary population for blue stragglers in NGC 188 (lower figure). (Figure 2 from {MG09}, reproduced with permission)}
\label{mdavies_figure9}
%%%
%similar to {matfig2} of Mathieu & Geller chapter!!!
%%%
%\end{minipage}
\end{figure}

By definition, in the case considered here, the donor is
the primary in the binary and will thus be more massive than the secondary
(mass-receiving) star. However, for systems with an initial mass ratio\index{mass ratio} close to unity,
it could be that in some systems an initial phase of mass transfer could change
the mass ratio such that the donor is now the less massive star in the binary.
What follows is then a phase of stable mass transfer where the envelope
of the primary is slowly transferred to the secondary star (on stellar evolutionary\index{stellar evolution}
timescales) whilst the separation of the binary increases.

Providing mass transfer occurs on the giant branch\index{red giant}, one would expect
to see at the end of this mass transfer a rejuvenated secondary star (now
the more massive star in the system) orbiting around some form of white dwarf\index{white dwarf}
(the former core of the primary). The binary separation being a few times larger
than red giant radii.

In Fig.~ \ref{mdavies_figure9}\footnote{This is a repetition of Fig. 3.2, reproduced here for the convenience of the reader.} we plot the binary properties of both 
solar-type main sequence stars\index{main sequence star} and blue stragglers observed in the open
cluster NGC 188\index{NGC 188} (\cite{MG09}; see also Chap. 3).  Interestingly, the vast majority of
the blue stragglers are in relatively {\it wide} binaries. This strongly  suggests 
that these systems have passed through a period of stable mass transfer from
a star evolving up the red-giant branch as described above. This is also
consistent with the observations in the field that blue stragglers
are in wide binaries \cite{Carney01,PS00, SPC03}.

\section{Comparing Primordial and Collisional Formation Rates in Clusters}
\label{davsec:6}

We may compute how the blue straggler production rate scales with cluster
mass, assuming all blue stragglers are made via the two-body collisions\index{two-body collision} described
above, and that these collisions\index{collision} occur exclusively within the dense core.

\begin{figure}
%\centering
%\resizebox{11truecm}{!}{\includegraphics{isothermal.eps}}
%\resizebox{13truecm}{!}{
\includegraphics[width=119mm]{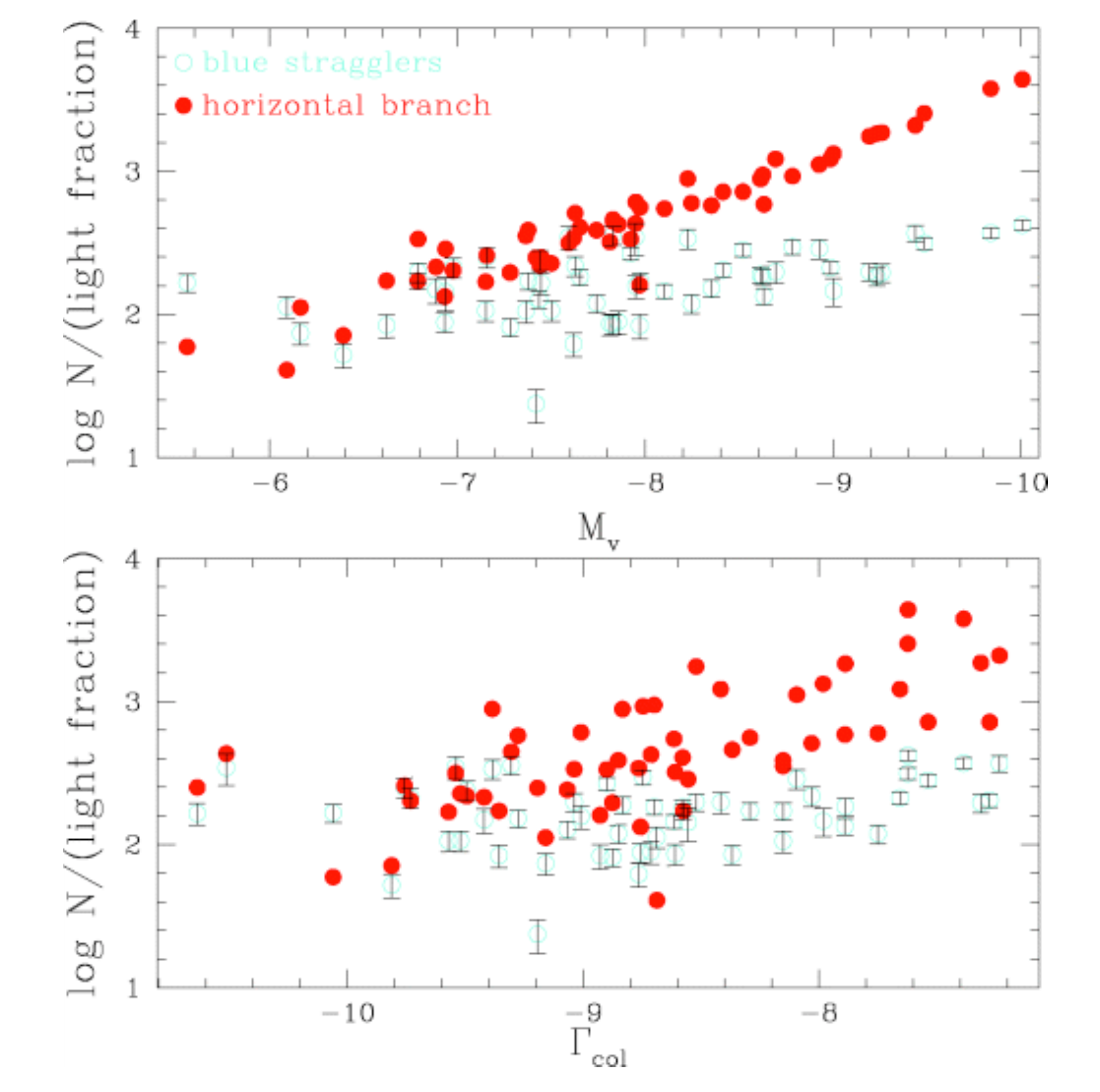}
%\begin{minipage}{13truecm}
\caption{The estimated total number of blue stragglers and horizontal
branch stars in the sample of 56 globular clusters as a function of cluster
total magnitude $M_{\rm V}$ (top panel) and stellar collision rate (bottom panel).
See \cite{Piotto04} for more details.
(Figure 1 from \cite{davDPd04}, reproduced with permission)}
\label{mdavies_figure10}
%\end{minipage}
\end{figure}

The stellar collision rate within the cluster core is given by
$\Gamma_{\rm coll} \propto {\rho^2 r_{\rm c}^3 / \sigma}$, where
$\rho$ is the mass density of stars within the cluster core, $r_{\rm
c}$ is the core radius, and $\sigma$ is the velocity dispersion of the
stars which is $\propto \sqrt{M_{\rm tot}/r_{\rm h}}$, where 
$M_{\rm tot}$ is the cluster total mass and $r_{\rm h}$ is
the radius containing half of the cluster's total mass.
 Also the cluster's  core mass $M_{\rm c}
\propto \rho r_{\rm c}^3$. Hence we have

\begin{equation}
\Gamma_{\rm coll} \propto {\rho^2 r_{\rm c}^3 \over \sigma}
\propto {\rho^2 r_{\rm c}^3 \over \sqrt{M_{\rm tot}/r_{\rm h}} }
\propto {M_{\rm c}^2 r_{\rm c}^{-3} \over \sqrt{M_{\rm tot}/r_{\rm h}} }
\propto { f_{\rm c}^2 r_{\rm h}^{1/2} \over r_{\rm c}^3 } M_{\rm tot}^{3/2}\ ,
\end{equation}
where $f_{\rm c} = M_{\rm c} / M_{\rm tot}$. Assuming for simplicity
$f_{\rm c}$, $r_{\rm c}$, and $r_{\rm h}$ are the same for all
clusters, we see that $\Gamma_{\rm coll} \propto M_{\rm tot}^{1.5}$.
Clearly $f_{\rm c}$, $r_{\rm c}$, and $r_{\rm h}$ all vary
between clusters, though this simply produces a spread around the
relationship. In other words, if all the blue stragglers in globular clusters
 really were produced via two-body collisions, then we would 
 expect to see that the number of blue stragglers increases
 with cluster mass as

\begin{equation}
  N_{\rm bs, coll} \propto M_{\rm tot}^{1.5}\ ,
  \label{mdavies_equation6}
\end{equation}
 This is not seen in the observed systems as illustrated 
 in  Fig.~ \ref{mdavies_figure10} where we see that the number of blue
 stragglers is relatively independent of cluster mass
 (see also \cite{Piotto04}).

We consider now the production of blue stragglers via binary evolution\index{binary evolution}, either
through the merger\index{merger} of the two components of a binary, or via mass transfer\index{mass transfer} 
from the  primary to the secondary as the primary evolves off the main sequence\index{main sequence}.
In order for such a process to produce the blue stragglers we see today, the mass
of the merger product,  or mass transfer enhanced secondary, must exceed
the current turn-off mass. The merger or 
mass transfer event must also have occurred relatively recently, i.e. less than the
blue straggler lifetime ago. If we assume here that the merger or mass transfer
is driven by the evolution of the primary off the main sequence, then this is 
equivalent to requiring that the primary mass subtracted by the current cluster
turn-off\index{turn-off} mass is less than some amount. For example, if we take a blue straggler
lifetime of one gigayear, and a turn-off mass today of $0.8$ M$_{\rm \odot}$, then
we require that the primary mass is in the range $0.8$ M$_{\rm \odot} \leq
M_1 \leq 0.816 $ M$_{\rm \odot}$. We note that this is a rather narrow range
of masses (due to the very strong mass dependance of main sequence lifetimes).
Typically, only a small fraction of binaries will satisfy this condition. If we consider
a binary population where stars are drawn from a reasonable Initial Mass Function (IMF; in our case, from \cite{EFT89}),
then we find that the fraction of binaries\index{binary fraction} making a blue straggler seen today
would be $f_{\rm bs} \simeq 0.006$. If the binaries
which produce blue stragglers are allowed to evolve in globular clusters\index{globular cluster}
without any interactions with other stars, then we would simply expect
that the number of blue stragglers derived from these primordial binaries
would be proportional to the cluster mass:

\begin{equation}
N_{\rm bs,bin} \propto f_{\rm bs,bin} M_{\rm tot}\,
\label{mdavies_equation7}
\end{equation}
where $f_{\rm bs,bin}$ is the fraction of the original binary population contributing to the 
blue straggler population today, i.e. those with primaries in the mass range
$0.8$ M$_{\rm \odot} \leq M_1 \leq 0.816 $ M$_{\rm \odot}$\ .

Let us assume that the two mechanisms described above are the only two
contributing to the observed blue straggler population seen in globular clusters.
How could their combination produce a population which is relatively independent
of cluster mass, given the mass dependencies described in Eq.~ (\ref{mdavies_equation6}) and (\ref{mdavies_equation7})?

The key point here is that dynamical interactions\index{dynamical interaction} occurring within the stellar cluster\index{stellar cluster}
which will alter the binary population: 1) exchange encounters with single stars occur
where less-massive stars in binaries are replaced by more-massive single stars;
2) stellar collisions may occur during binary-single encounters (see Figure
\ref{mdavies_figure5}) which lead to mergers and may remove the binary
from the population; 3)  some binaries may be destroyed by binary-binary encounters.

\begin{figure}[ht]
\sidecaption
%\centering
%\resizebox{11truecm}{!}{\includegraphics{isothermal.eps}}
%\resizebox{10truecm}{!}{
\includegraphics[width=75mm]{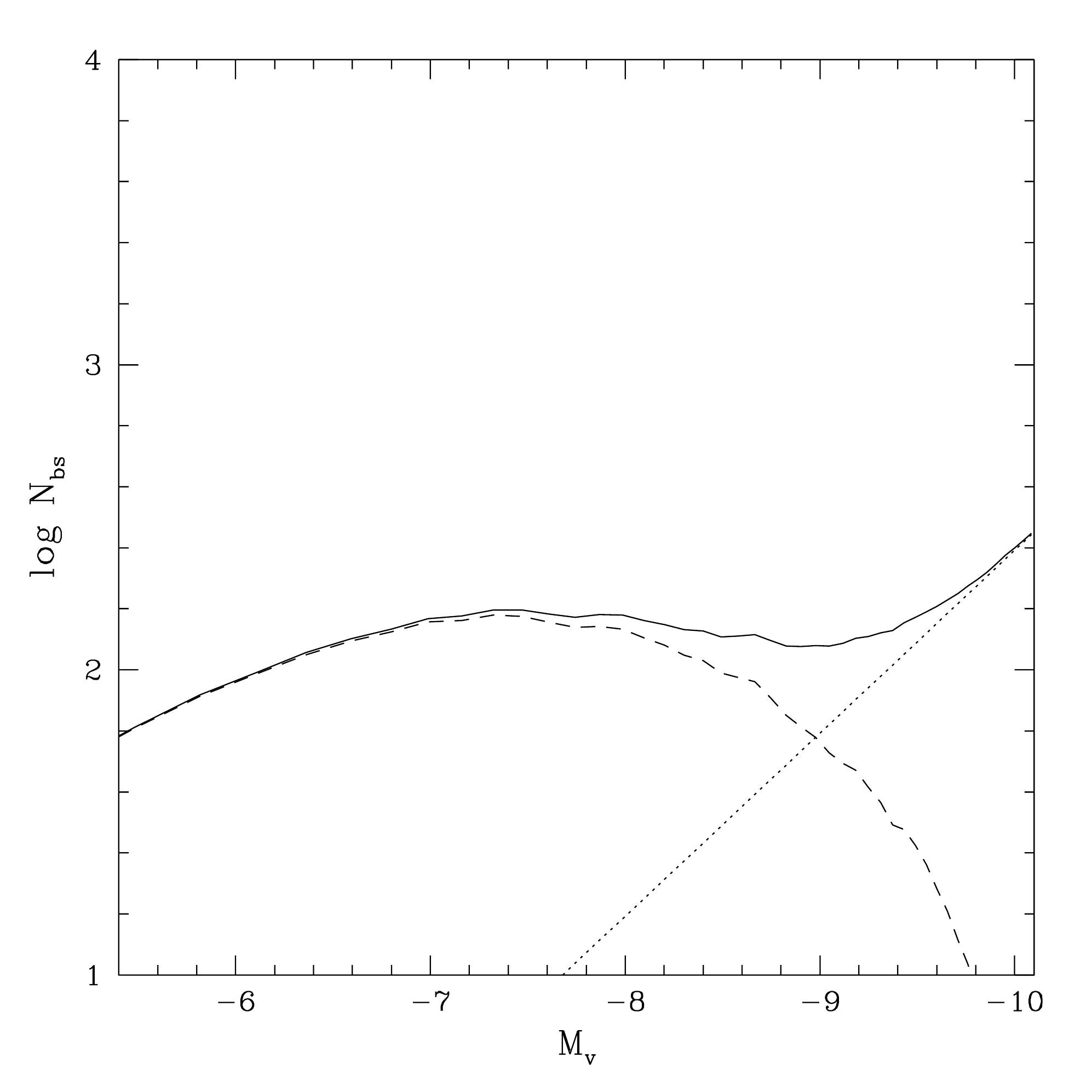}
%\begin{minipage}{13truecm}
\caption{The number of BSSs produced over the last Gyr as a function of absolute cluster 
luminosity\index{luminosity}, $M_V$ , assuming $M/L_V = 3$ for all clusters\index{cluster}. The contribution from primordial
 systems is shown with a dashed line, whilst those produced via collisions\index{collision} (involving either 
 two single stars or binaries) is shown as a dotted line. The total is given as a solid line.
 (Figure 6 from \cite{davDPd04}, reproduced with permission)}
\label{mdavies_figure11}
%\end{minipage}
\end{figure}

Thus in more massive clusters,
the binaries will on average have experienced more close
encounters with single stars and binaries.
Binary-single encounters scale with cluster mass in the same way as
two-body collisions seen in equation (\ref{mdavies_equation6}). 
From exchange encounters, the binaries will tend to contain
  more massive stars
today. This will {\it reduce}
the fraction of binaries contributing to the blue straggler population today, 
as more of the primaries in the binaries will have evolved off the main sequence
too long ago in the past: the  blue stragglers they produced have been and gone
by today \cite{davDPd04}. However, as pointed
out by Knigge (private communication; see also Chap. 13),
stellar evolution may reduce this  effect as stars will have evolved, becoming
(less-massive) compact remnants before they encounter binaries. 
If the binaries are in mass-segregated stellar cluster cores, more of the stars they encounter
will be massive and thus more massive stars will exchange into binaries before they
evolve. Even if this effect is limited, binary destruction via stellar collisions 
and binary-binary encounters may be equally effective in reducing the blue-straggler
population derived from binaries. The destruction of binaries via these two 
mechanisms is shown in Figure \ref{mdavies_figure7}
(see also \cite{HMR92}).
 Observational evidence for lower binary fractions
in more massive clusters (and thus, on average those having more dynamical
interactions) has been reported by  \cite{Milone12}.

\begin{figure}
\includegraphics[width=119mm]{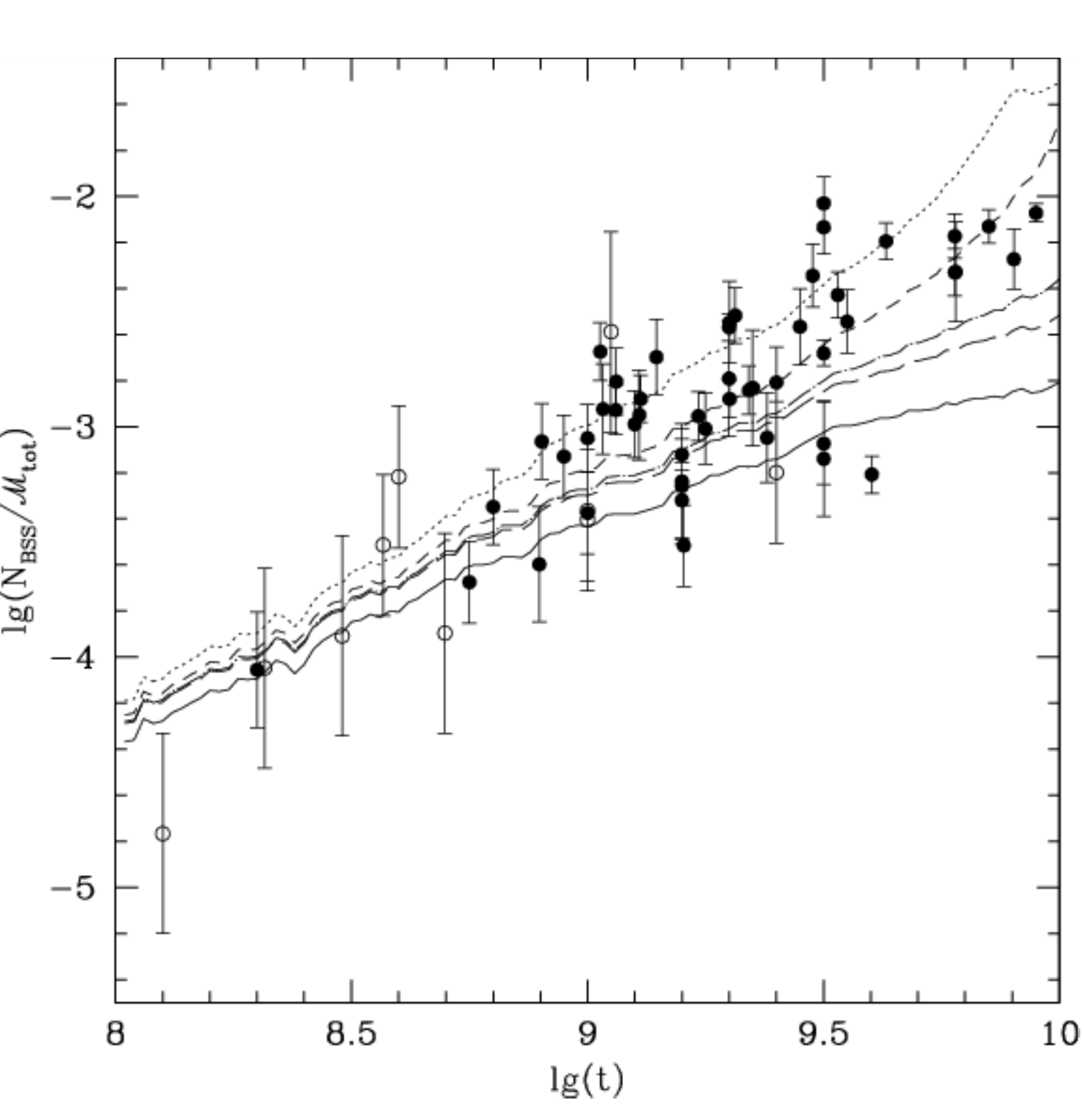}
%\begin{minipage}{13truecm}
\caption{Comparison between the expected number of blue stragglers
and the number observed in open clusters\index{open cluster}. The solid line represents
the model given in \cite{dM06} where it is assumed that the observed
blue stragglers are derived from primordial binaries which have undergone
mass-transfer. Dotted, dashed, dot-dashed and long-dashed curves reproduce
respectively the expected values obtained taking into account the evaporation
of stars from clusters following the models of \cite{TF05} with
2k, 8k, 32k, and 131k N-body models.
(Figure 8  from \cite{dM06}, reproduced with permission)}
\label{mdavies_figure12}
%\end{minipage}
\end{figure}

Combining the contribution from stellar collisions and that from binaries,
we have

\begin{equation}
N_{\rm bs}= k_{\rm bs,coll} M_{\rm tot}^{3/2} + 
k_{\rm bs,bin} f_{\rm bs,bin}(M_{\rm tot}) M_{\rm tot}\ ,
\end{equation}
where $k_{\rm bs,coll}$ and $k_{\rm bs,bin}$ are suitably chosen constants.
$f_{\rm bs,bin}(M_{\rm tot})$ has been  determined through Monte Carlo
calculations of binary-single encounters
including only the effects of exchange encounters, not including the
effects of stellar evolution \cite{davDPd04}.

The number of blue stragglers expected as a function of
absolute cluster magnitude\index{absolute magnitude} is shown in  Fig.~ \ref{mdavies_figure11}. 
Here we have assumed a mass-to-light ratio $M/L_V = 3$, 
and have  taken reasonable values for the two constants (see \cite{davDPd04}). 
Including the effects of stellar evolution could
reduce the effect due to exchanges, but the binary population will also be reduced
via stellar collisions during binary-single encounters and by binary destruction
during binary-binary encounters. The net effect is likely to be the same:
in environments where
interactions with other stars and binaries are sufficiently frequent,  the contribution to
the blue straggler population made by binaries is reduced. We see that
the blue straggler population derived from binaries dominates for most clusters
and that direct collisions\index{collision} only become important for clusters brighter than $M_{\rm v}
= -9 $ (or equivalently a mass, $M_{\rm tot} = 10^6 $ M$_{\odot}$). 
Indeed, by considering the number of blue stragglers found in cluster
cores and comparing this to the total stellar mass contained in cluster cores,
Knigge et al. \cite{KLS09} concluded that most blue stragglers come from binary systems ---
but see also \cite{LKS13}.

It is important
to recall that the trend shown in Fig.~ \ref{mdavies_figure11}
has been derived assuming average cluster properties
(the dependence on cluster mass given in Eq.~(\ref{mdavies_equation6})
 assumed for example
that all clusters have the same core and half-mass radii\index{half-mass radius}). There will be outliers
to the distribution shown here. Nonetheless, the turn over in the blue straggler
population derived from primordial binaries seen here due to encounters\index{encounter} with 
single stars (and perhaps also binary destruction through stellar collisions and binary-binary encounters\index{collision})
 does explain the observed, relatively flat, blue straggler population. 
There is observational evidence that two formation channels for blue stragglers occur
in at least one globular cluster --- M30\index{M30} ---  as two distinct blue-straggler 
sequences have been observed \cite{Ferraro09}.

It should also be noted that the {\it specific} frequency of blue stragglers
seen in clusters\index{cluster} (i.e. the number per unit mass) is in fact {\it less} than one 
would obtain for binaries in the field. This is because binary-single encounters 
(and binary-binary encounters) 
act to reduce the fraction of binaries contributing to the blue straggler population
today.  Thus the largest specific frequency of blue stragglers derived
from binaries will occur in low-density environments where no such
encounters are expected: in the low-density haloes of clusters and in 
field of the Galaxy\index{Galaxy}. Indeed, this is consistent with observations of blue
stragglers seen in the Galactic halo\index{halo} \cite{Carney01,PS00,SPC03}
and also in open clusters\index{open cluster},
where the observed population follows that expected if it is derived from primordial
binaries which have undergone mass transfer\index{mass transfer}, as shown in Fig.~ \ref{mdavies_figure12} \cite{dM06}.

\begin{acknowledgement}
This book is the result of a meeting held at ESO, Chile. I thank 
the local organisers for their hospitality.
I  thank Christian Knigge for pointing out the importance
of stellar evolution when considering the evolution of a population
of binaries within a stellar cluster. 
\end{acknowledgement}

\backmatter%%%%%%%%%%%%%%%%%%%%%%%%%%%%%%%%%%%%%%%%%%%%%%%%%%%%%%%
%\appendix
%\include{appendix}
%\include{glossary}
\printindex

%%%%%%%%%%%%%%%%%%%%%%%%%%%%%%%%%%%%%%%%%%%%%%%%%%%%%%%%%%%%%%%%%%%%%%


\begin{thebibliography}{99}

\bibitem{Bacon96}
Bacon, D., Sigurdsson, S.,  Davies, M. B.:  MNRAS {\bf 281}, 830 (1996)

\bibitem{davBH87}
Benz, W. ,  Hills, J. G.:  ApJ {\bf 323}, 614 (1987)

\bibitem{davBH92}
Benz, W. ,  Hills, J. G.:  ApJ {\bf 389}, 546 (1992)

\bibitem{Carney01}
Carney, B. W., Latham, D. W., Laird, J. B., Grant, C. E.,  Morse, J. A.: AJ {\bf 122}, 3419 (2001)

\bibitem{DBH91}
Davies, M. B., Benz, W.,  Hills, J. G.: ApJ {\bf 381}, 449 (1991)

\bibitem{DBH93}
Davies, M. B., Benz, W.,  Hills, J. G.: ApJ {\bf 411}, 285 (1993)

\bibitem{DBH94}
Davies, M. B., Benz, W.,  Hills, J. G.:  ApJ {\bf 424}, 870 (1994)

\bibitem{DH98}
Davies, M. B. ,  Hansen, B. M. S.: MNRAS {\bf 301}, 15 (1998)

\bibitem{davDPd04}
Davies, M. B., Piotto, G.,  de Angeli, F.: MNRAS {\bf 349}, 129 (2004)

\bibitem{dM06}
de Marchi, F., de Angeli, F., Piotto, G., Carraro, G.,  Davies, M. B.: A\&A {\bf 459}, 489 (2006)

\bibitem{EFT89}
Eggleton, P. P., Fitchett, M. J.,  Tout, C. A.: ApJ {\bf 347}, 998 (1989)

\bibitem{davFPR75}
Fabian, A. C., Pringle, J. E.,  Rees, M. J.:  MNRAS {\bf 172}, 15P (1975)

\bibitem{Ferraro09}
Ferraro, F. R., Beccari, G., Dalessandro, E., et al.:  Nature {\bf 462}, 1028 (2009)

\bibitem{GP08}
Glebbeek, E. ,  Pols, O. R.: A\&A {\bf 488}, 1017 (2008)

\bibitem{GPH08}
Glebbeek, E., Pols, O. R.,  Hurley, J. R.: A\&A {\bf 488}, 1007 (2008)

\bibitem{Heggie75}
Heggie, D. C.: MNRAS {\bf 173}, 729 (1975)

\bibitem{HMG92}
Hut, P., McMillan, S., Goodman, J., et al.: PASP {\bf 104}, 981 (1992)

\bibitem{HMR92}
Hut, P., McMillan, S.,  Romani, R. W.: ApJ {\bf 389}, 527 (1992)

\bibitem{HV83}
Hut, P. ,  Verbunt, F.: Nature {\bf 301}, 587 (1983)

\bibitem{KLS09}
Knigge, C., Leigh, N.,  Sills, A.: Nature {\bf 457}, 288 (2009)

\bibitem{LKS13}
Leigh, N., Knigge, C., Sills, A., et al.: MNRAS {\bf 428}, 897 (2013)

\bibitem{davLeonard89}
Leonard, P. J. T.: AJ {\bf 98}, 217 (1989)

\bibitem{LRS95}
Lombardi, Jr., J. C., Rasio, F. A.,  Shapiro, S. L.: ApJ {\bf 445}, L117 (1995)

\bibitem{LRS96}
Lombardi, Jr., J. C., Rasio, F. A.,  Shapiro, S. L.: ApJ {\bf 468}, 797(1996)

\bibitem{Lom02}
Lombardi, Jr., J. C., Warren, J. S., Rasio, F. A., Sills, A.,  Warren, A. R.: ApJ {\bf 568}, 939 (2002)

\bibitem{MG09}
Mathieu, R. D. ,  Geller, A. M.: Nature {\bf 462}, 1032 (2009)

\bibitem{davMcCrea64}
McCrea, W. H.: MNRAS {\bf 128}, 147 (1964)

\bibitem{Milone12}
Milone, A. P., Piotto, G., Bedin, L. R., et al.: A\&A {\bf 540}, A16 (2012)

\bibitem{PFa09}
Perets, H. B. ,  Fabrycky, D. C.: ApJ {\bf 697}, 1048 (2009)

\bibitem{Piotto04}
Piotto, G., De Angeli, F., King, I. R., et al.: ApJ {\bf 604}, L109 (2004)

\bibitem{PS00}
Preston, G. W. ,  Sneden, C.: AJ {\bf 120}, 1014 (2000)

\bibitem{SanBH97}
Sandquist, E. L., Bolte, M.,  Hernquist, L.: ApJ {\bf 477}, 335 (1997)

\bibitem{SP93}
Sigurdsson, S. ,  Phinney, E. S.: ApJ {\bf 415}, 631(1993)

\bibitem{SAD05}
Sills, A., Adams, T.,  Davies, M. B.: MNRAS {\bf 358}, 716 (2005)

\bibitem{SADB02}
Sills, A., Adams, T., Davies, M. B.,  Bate, M. R.: MNRAS {\bf 332}, 49 (2002)

\bibitem{davSills01}
Sills, A., Faber, J. A., Lombardi, Jr., J. C., Rasio, F. A.,  Warren, A. R.: ApJ {\bf 548}, 323 (2001)

\bibitem{SGCR13}
Sills, A., Glebbeek, E., Chatterjee, S.,  Rasio, F. A.: ApJ {\bf 777}, 105  (2013)

\bibitem{SKL09}
Sills, A., Karakas, A.,  Lattanzio, J.: ApJ {\bf 692}, 1411 (2009)

\bibitem{davSills97}
Sills, A., Lombardi, Jr., J. C., Bailyn, C. D., et al.: ApJ {\bf 487}, 290 (1997)

\bibitem{SPC03}
Sneden, C., Preston, G. W.,  Cowan, J. J.: ApJ {\bf 592}, 504 (2003)

\bibitem{TF05}
Tanikawa, A. ,  Fukushige, T.: PASJ {\bf 57}, 155
(2005)

\bibitem{Vilhu82}
Vilhu, O.: A\&A {\bf 109}, 17 (1982)

\end{thebibliography}
\end{document}